\documentclass[twocolumn,twoside,slac_two]{revtex4}
\usepackage{graphicx}
\usepackage{fancyhdr}
\usepackage{amsmath,amssymb}

\newcommand{\E}{E_{0}}
\newcommand{\xmax}{x_{max}}

\begin{document}

\title{The depth of maximum shower development and its fluctuations: cosmic ray mass composition at $E_{0} \ge 10^{17}$~eV}

\author{S.~P.~Knurenko}
\author{A.~Sabourov}
\affiliation{Yu. G. Shafer Institute of cosmophysical research and aeronomy}

\begin{abstract}
  We present new data on Cherenkov light observations obtained during the period 1994-2009, after a modernization of the Yakutsk EAS array. A complex analysis of $\xmax$ and its fluctuations $\sigma(\xmax)$ was performed over a wide energy range. With the new data, according to QGSJet~II model, an estimation was made of the cosmic ray mass composition for $\E \sim 10^{17} - 3 \times 10^{19}$~eV. The result points towards a mixed composition with a large portion of heavy nuclei at $\E \sim 10^{17}$~eV and the dominance of light nuclei at $\E \sim 10^{19}$~eV. The analysis of $\sigma(\xmax)$ energy dependence for the same energies qualitatively confirms this result. The shape of the $\xmax$ distribution at fixed energy $10^{18}$~eV is analysed to make more precise conclusions on cosmic ray mass composition.
\end{abstract}

\maketitle
\thispagestyle{fancy}

\section{Introduction}

The Yakutsk EAS array effectively covers the energy domain from $10^{15}$~eV to $5 \times 10^{20}$~eV by measuring charged particles, muons with $\varepsilon_{\text{thr}} \ge 1$~GeV$\cdot \sec{\theta}$ and Cherenkov light emission. In the same energy region two irregularities of cosmic ray (CR) energy spectrum are observed~--- the {\em knee} ($3 \times 10^{15}$~eV) and the {\em ankle} ($8 \times 10^{18}$~eV). The nature of these irregularities is yet to be explained. From the work by~\citet{bib1} it follows that such a spectrum shape could be a consequence of a transition from galactic to extragalactic components in the total CR flux, i.e. there must be some region in the CR spectrum where the intensities of these fluxes become equal to each other and then decrease. The boundaries of such a transition region is yet unknown and represents a subject of research at many large EAS arrays. The physics of this phenomenon is tightly connected with particle drift in the magnetic fields of our Galaxy and outer space. Since nuclei of different masses behave differently in a magnetic field, then the CR composition in different points of space will differ. Therefore, it is possible to solve the problem of the transitional region by measuring the CR mass composition in EAS related experiments.

It is a known fact that the depth of shower maximum ($\xmax$) and fluctuations in EAS development are sensitive to the atomic number of the primary particle and for this reason they are used to estimate the CR mass composition~\citep{bib2, bib3, bib4}. It is especially important for the ultra-high energy region where direct measurements of mass composition are impracticable. For example, works by \citet{bib5, bib6, bib7} utilized single characteristics and their combinations: $\left<\xmax\right>$, $\sigma(\xmax)$ and $d\xmax / d\lg{\E}$. These works provided initial estimations of inelastic interaction cross-sections at ultra-high energies.

In this paper we present the data on longitudinal EAS development reconstructed from Cherenkov emission data. These data were obtained after modernization of the Yakutsk array when the accuracy of the main EAS characteristics increased  compared to the previous series of observations. A precise knowledge of the mass composition together with the energy spectrum plays a major role in understanding CR astrophysics~\citep{bib8}. In this sense, engaging the maximal possible number of composition-sensitive EAS characteristics increases the reliability of CR chemical composition estimation. It is important to consider not only the mean EAS parameters, e.g $\xmax$, muon content $\rho_{\mu}/\rho_{\text{ch}}$ but also their fluctuations in given energy intervals~\citep{bib9, bib10}. In order to minimize the latter, it is also a good idea to analyze them at fixed energies.

\section{Technical aspects of longitudinal EAS development characteristics estimation}

The determination of $\xmax$ in individual showers is based on methods developed at the Yakutsk array and utilize the measurements of EAS Cherenkov light emission at different core distances. In the first method,  $\xmax$ is determined by the parameter $p = \lg{Q_{200} / Q_{550}}$ (a relation of Cherenkov light fluxes at $200$ and $550$~m from the core); the second ~--- involves the reconstruction of the EAS development cascade curve, using the Cherenkov light lateral distribution function and a reverse solving~\citep{bib11}; the third is based on half-width and half-height of Cherenkov light pulses recorded at $200$~m from the core; the fourth method includes recording the Cherenkov track with several detectors based on camera-obscura located at $300-500$~m from the array center~\citep{bib12}. %Examples demonstrating these techniques for $\xmax$ estimation are shown on figures~\ref{fig1a}, \ref{fig1b} and \ref{fig1c}.

%\begin{figure}
%  \centering
%  \includegraphics[width=0.45\textwidth, clip]{fig1a}
%  \caption{Estimation of $\xmax$ by parameter $p = \lg{Q_{200} / Q_{550}}$}
%  \label{fig1a}
%\end{figure}
%
%\begin{figure}
%  \centering
%  \includegraphics[width=0.35\textwidth, clip]{fig1b}
%  \caption{Estimation of $\xmax$ by half-width and half-height of Cherenkov %pulse $\tau_{1/2}$}
%  \label{fig1b}
%\end{figure}
%
%\begin{figure}
%  \centering
%  \includegraphics[width=0.35\textwidth, clip]{fig1c}
%  \caption{$\xmax$ estimation by signals from different atmospheric depths recorded with tracking Cherenkov detectors}
%  \label{fig1c}
%\end{figure}

Various factors affect the methods mentioned above: the way the showers are selected, precision of core location, atmosphere transparency, mathematical methods used to calculate parameters of approximated functions, hardware-related fluctuations and so on. The influence from single and composite factors on physical results of the Yakutsk array operation was calculated either with full simulation of the measurement procedure or empirically estimated during special methodical experiments. For instance, an estimation of hardware-related errors was performed by the analysis of two nearby detectors that measure charged particles, muons and Cherenkov light emission~\citep{bib6, bib13}. The accuracy of $\xmax$ determination in individual showers was estimated in simulating EAS characteristics measurements at the array involving Monte-Carlo methods and amounted to $30-45$~g/cm$^{2}$, $35-55$~g/cm$^{2}$, $15-25$~g/cm$^{2}$, $35-55$~g/cm$^{2}$ respectively for the first, second, third and fourth methods. The total error of $\xmax$ estimation included errors associated with core location, atmospheric transparency during the observational period, hardware fluctuations and mathematical methods used to calculate main parameters.

\section{Mean depth of maximum shower development}

\begin{figure}
  \centering
  \includegraphics[width=0.45\textwidth, clip]{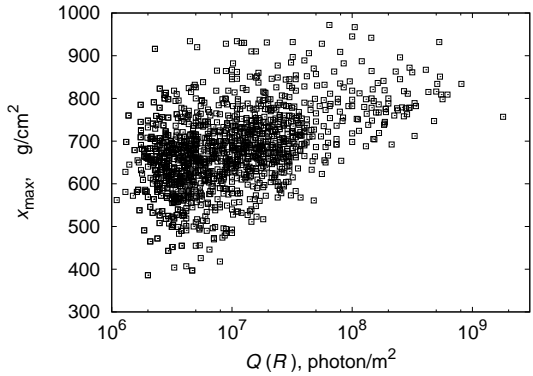}
  \caption{Individual EAS events detected in the Yakutsk experiment during 1994 -- 2009.}
  \label{fig2}
\end{figure}

\begin{figure}
  \centering
  \includegraphics[width=0.45\textwidth, clip]{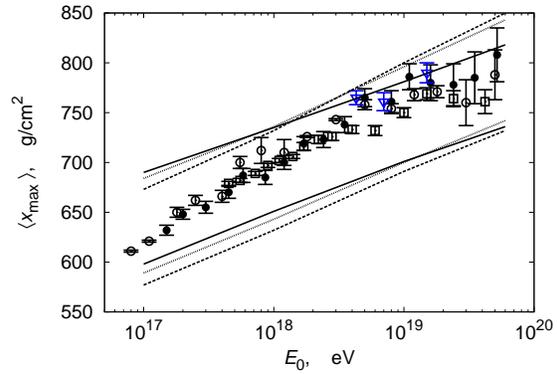}
  \caption{Energy dependence of $\xmax$. Filled circles represent the Yakutsk data, open circles~--- CASA-MIA, squares~--- AUGER data, blue triangles~--- preliminary results of the Telescope Array experiment \citep{TA}. Solid lines~--- results obtained with QGSJet~II, dashed~--- EPOS~1.6, point line~--- SIBYLL~1.62}
  \label{fig3}
\end{figure}

Figure~\ref{fig2} demonstrates a cloud of points in the $\xmax$ distribution for showers with energy above $10^{17}$~eV. These data were obtained using all four methods and reflect an alteration of $\xmax$ towards lower atmosphere depths with increasing energy. Figure~\ref{fig3} shows $\xmax$ values averaged over energy intervals together with the data from other experiments. On the same picture results of different hadron models calculations are shown. All experimental data coincide within experimental errors and demonstrate an irregular shift with energy. Up to $3 \times 10^{18}$~eV E.R. has value $60-80$~g/cm$^{2}$ and within the interval of $3 \times 10^{18} - 5 \times 10^{19}$~eV it equals  $40-60$~g/cm$^{2}$. This might be interpreted as a possible alteration in mass composition at very high energies. A comparison with calculations reveals a tendency for a light nuclei abundance starting from $5 \times 10^{17}$~eV to $2 \times 10^{19}$~eV and some abundance above $2 \times 10^{19}$~eV.

\section{Fluctuations of $\xmax$}

\begin{figure}
  \centering
  \includegraphics[width=0.45\textwidth, clip]{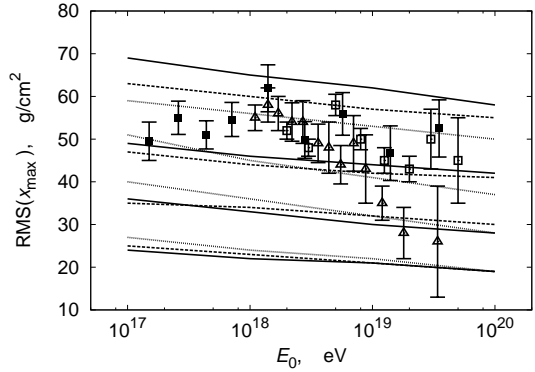}
  \caption{Fluctuations of the depth of maximum EAS development: filled squares~--- Yakutsk data, open squares~--- HiRes data, open triangles~--- data from Pierre Auger Observatory. Straight line~--- results obtained with QGSJet01, dashed line~--- QGSJet~II, dotted line~--- SIBYLL~1.62 for various primary nuclei (see~\citet{bib15})}
  \label{fig4}
\end{figure}

\begin{figure}
  \centering
  \includegraphics[width=0.45\textwidth, clip]{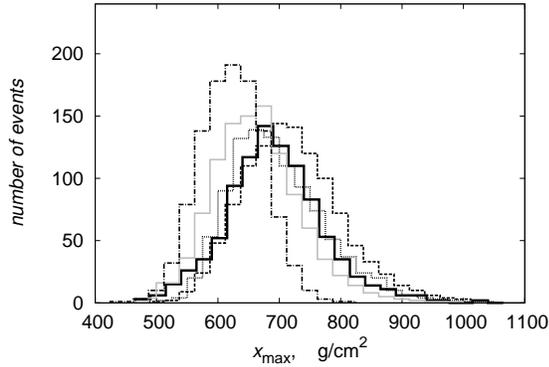}
  \caption{$\xmax$ distribution at fixed energy $10^{18}$~eV. Solid line represents the Yakutsk data ($8 \times 10^{17} < \E < 2 \times 10^{18}$~eV, $\left<\E\right> = 1.0 \times 10^{18}$~eV, $857$~events); dotted line~--- QGSJet01 for mixed composition ($70$\,\%~p, $30$\,\%~Fe); dashed line~--- QGSJet01 for primary protons, solid grey line~--- QGSJet01 for CNO group nuclei, dash-dotted line~--- QGSJet01 for iron nuclei (see~\citet{bib4})}
  \label{fig5}
\end{figure}

Fluctuations of $\xmax$ play a huge role in EAS longitudinal development as they are associated with the point of first interaction (and, hence, with  the  inelastic interaction cross-section, $\sigma_{\text{A-air}}$), energy transfer to secondary hadron particles (inelastic coefficient $K_{\text{inel}}$) and, to a great extent, depend on the kind of primary particle initiating a shower. So, the amount of fluctuations measured in different energy intervals could characterize the CR mass composition at a given energy and on the whole determine the dynamics of its change with the energy of the primary particle. Figure~\ref{fig4} demonstrates the energy dependence of $\sigma(\xmax)$ obtained at the Yakutsk array and for comparison the same figure shows the HiRes data~\citep{bib15}. The data from the HiRes experiment virtually reproduces the data from Yakutsk but have a slight tendency of $\sigma(\xmax)$ change: a small increase in the region of $10^{17} - 10^{18}$~eV and a decrease at $2 \times 10^{18} - 5 \times 10^{19}$~eV. The curves representing simulation results, obtained with QGSJet01, QGSJet~II and SIBYLL models, are also shown on this figure. Calculations were performed for proton, helium nuclei, CNO group and iron nuclei. Comparison with experimental data has shown that the CR composition in this energy region is mixed with a dominance of protons and helium nuclei. It should be pointed out that according to Figure~\ref{fig4}, the portion of heavy nuclei in the CR flux energy above $2 \times 10^{18}$~eV is small and helium and CNO-group nuclei might play a significant role. We came to the same conclusion \citep{bib4} where the shape of the $\xmax$ distribution was analyzed within the framework of the QGSJet01 model at fixed energies $10^{18}$~eV and $\sim 10^{19}$~eV (see Figure~\ref{fig5}).

\section{Cosmic ray mass composition}

%Experimental data from three EAS experiments represented on Figure~\ref{fig6} reveal some tendency of change in $\left<\ln{A}\right>$ value at energy $\sim 10^{19}$~eV. While in the interval $10^{17} - 10^{18}$~eV a decrease of $\left<\ln{A}\right>$ to values $1.3-1.5$ took place, at energies above $10^{19}$~eV a slight growth is observed. However, in this energy domain there is still significant discrepancy in the data due to poor statistics. This is why the reliability of our statement is quite limited.

\begin{figure}[t!]
  \centering
  \includegraphics[width=0.45\textwidth, clip]{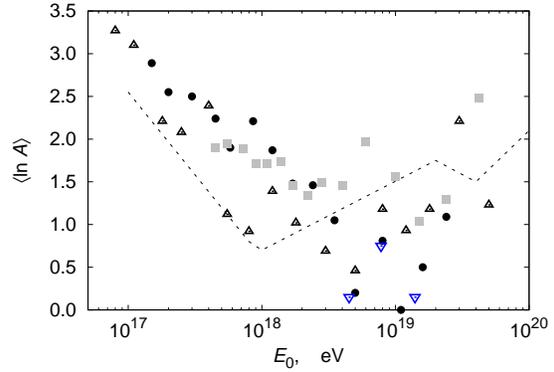}
  \caption{Mean mass number of primary particle as a function of energy. Circles represent the Yakutsk data, triangles~--- HiRes data, squares~--- results obtained at Auger observatory, blue empty triangles~--- preliminary data from the Telescope Array experiment \cite{TA}, dotted line~--- computational results by \citet{bib17}}
  \label{fig6}
\end{figure}

Figure~\ref{fig6} displays the mean natural logarithm of the CR atomic number $\left<\ln{A}\right>$ concluded from the $\xmax$ data from four experiments~--- Yakutsk, HiRes, Auger and Telescope Array \citep{TA}. For $\left<\ln{A}\right>$ derivation $\xmax$ values were utilized, obtained in simulations within the framework of the QGSJet~II models for proton and iron nuclei. The $\left<\ln{A}\right>$ value was calculated according to the relation proposed by \citet{bib16}:
\begin{equation}
  \left<\ln{A}\right> = \frac{\xmax - \xmax^{\text{H}}}{\xmax^{\text{Fe}} - \xmax^{\text{H}}} \cdot \ln{56}
  \label{eq:1}
\end{equation}

At first glance, all data reveal a tendency to change $\left<\ln{A}\right>$ with energy. For instance, in the energy interval $2 \times 10^{17} - 3 \times 10^{18}$~eV, the value of $\left<\ln{A}\right>$ drops from $3$ to $1.3$ and above $10^{18}$~eV a slight increase is noted. Such a behaviour is close to the ``dip''-scenario from the work by \citet{bib17}, where two peaks are seen in the energy dependence of $\left<\ln{A}\right>$. The first one, at $\sim 10^{17}$~eV, corresponds to the ending of the galactic component, the second~--- at $10^{19}$~--- to the start of CR intensity change due to GZK-cutoff.

However, there is still a significant data dispersion in this energy region due to poor event statistics. Thus, the reliability of our statement is quite limited. For a more precise conclusion on ultra-high energy cosmic rays origin, a few conditions must be fulfilled: improved statistics, improvement of $\xmax$ estimation precision, adaptation of a single hadron interaction model that well describes experimental data and involving several alternative methods for $\xmax$ evaluation.

\begin{table*}
  \caption{New data from the Yakutsk array}
  \vskip4mm
  \centering
  \begin{tabular}{cccccccccc}
    \hline
    $\E$, eV & $1.5 \times 10^{17}$ & $2.0 \times 10^{17}$ & $3.0 \times 10^{17}$ & $4.5 \times 10^{17}$ & $5.8 \times 10^{17}$ & $8.6 \times 10^{17}$ & $1.2 \times 10^{18}$ & $1.7 \times 10^{18}$ & $2.4 \times 10^{18}$ \\
    \hline
    $\xmax$ & $632$ & $648$ & $655$ & $670$ & $687$ & $685$ & $700$ & $719$ & $723$ \\
    \hline
    $\sigma(\xmax)$ & $5$ & $5$ & $6$ & $6$ & $7$ & $7$ & $7$ & $7$ & $8$ \\
    \hline
    \hline
    $\E$, eV & $3.5 \times 10^{18}$ & $5.0 \times 10^{18}$ & $8.0 \times 10^{18}$ & $1.1 \times 10^{19}$ & $1.6 \times 10^{19}$ & $2.4 \times 10^{19}$ & $3.5 \times 10^{19}$ & $5.1 \times 10^{19}$ & -- \\
    \hline
    $\xmax$ & $738$ & $765$ & $761$ & $786$ & $780$ & $778$ & $785$ & $808$ & -- \\
    \hline
    $\sigma(\xmax)$ & $8$ & $9$ & $11$ & $13$ & $18$ & $21$ & $26$ & $25$ & -- \\
    \hline
  \end{tabular}
\end{table*}

\section{Conclusions}

Thus, according to all the data reviewed above, within the framework of the QGSJet hadron interaction model it is reasonable to speculate that the primary cosmic ray mass composition alters during the energy transition from $10^{17}$~eV to $5 \times 10^{18}$~eV. At $\E \ge 5 \times 10^{18}$~eV  $\sim 70$\,\% of cosmic rays consist of protons and helium nuclei. The content of other nuclei in the region of the ankle of the spectrum does not exceed $\sim 30$\,\%. A large portion of protons and helium nuclei in the primary CR near the ankle is most likely associated with a significant contribution from particles arriving from outside our Galaxy. In such a case the region of transition from galactic to extragalactic component might be in the energy interval $10^{17} - 10^{19}$~eV. The problem of mass composition altering above $10^{19}$~eV remains unresolved due to poor event statistics.

\end{document}